\documentclass[english,aps,prl,amsmath,amssymb,twocolumn,letterpaper,citeautoscript,superscriptaddress,longbibliography]{revtex4-1}

\usepackage{mathtools}

\usepackage[normalem]{ulem} 

\usepackage{comment}
\usepackage{amsmath}
\usepackage{physics}
\usepackage{graphicx}
\usepackage{babel}
\usepackage{amstext}
\usepackage{esint}
\usepackage[bookmarks=false,linkcolor=blue,urlcolor=blue,colorlinks,citecolor=blue]{hyperref}
\usepackage{pdfpages}	
\usepackage{pgffor}

\makeatletter
\AtBeginDocument{\let\LS@rot\@undefined} 
\makeatother

\begin{document}

\title{Supercurrent as a bulk probe for topological phase} 

\author{Tatiana de Picoli}

\affiliation{Department of Physics and Astronomy, Purdue University, West Lafayette, Indiana 47907 USA}

 \author{Ian Wojtowicz}
\affiliation{Department of Physics and Astronomy, Purdue University, West Lafayette, Indiana 47907 USA}
\affiliation{Department of Mechanical Engineering, University of Minnesota Twin Cities, Minneapolis, 55455, USA}

\author{Jukka I. V\"ayrynen}

\affiliation{Department of Physics and Astronomy, Purdue University, West Lafayette, Indiana 47907 USA}

\date{\today}
\begin{abstract}

Distinguishing trivial from topological superconducting phase remains a central and contested experimental challenge, since most existing probes infer the bulk phase transition indirectly, from boundary 
signatures that can be  sensitive to local physics. 
Here, we propose an  alternative that probes the bulk topological phase and is based on  supercurrent response. 
We study a one-dimensional Rashba spin-orbit-coupled superconducting nanowire with an applied magnetic field that opens a topological gap. 
We show that for strong spin-orbit coupling, the superfluid stiffness, and consequently the supercurrent, is suppressed by a factor of two inside the topological phase, producing a non-monotonic signature of the bulk phase transition as a function of chemical potential. 
We further use a low-energy helical model to understand this suppression analytically. 
Our results show that the bulk topological phase transition can be detected through kinetic inductance measurements. 
\end{abstract}
\maketitle

\textbf{\emph{Introduction.~}}

The pursuit of fault-tolerant quantum computing has motivated numerous studies on topological phases of quantum matter~\cite{2003AnPhy.303....2K,RevModPhys.80.1083,qi_review,HasanReview,simonbook}. Topological superconductors can host exotic excitations at their boundaries, such as Majorana zero modes, which are one of the main candidates for topological quantum computation~\cite{Alicea2012,kitaev01,frolov2020topological,doi:10.1126/science.aav3392,flensberg2021engineered,aasen2025roadmapfaulttolerantquantum}. In this field, one central challenge remains distinguishing trivial from topological states in a realistic experimental setup~\cite{prada2020andreev}.

In the literature, various probes of topological phase transitions have been proposed, including approaches based on transport, spectroscopy, and thermodynamic properties~\cite{fu2009probing,lutchyn2010majorana,wimmer2011quantum, beenakker2013search,fregoso2013electrical,2014NatCo...5E4772G,PhysRevB.92.134508,ben2015detecting,mishmash2016approaching,szumniak2017spin,huang2018quasiparticle,chevallier2018topological,dartiailh2021phase,ren2024,bottcher2024circuit}. In one-dimensional Rashba nanowires with proximity-induced pairing and a Zeeman field~\cite{lutchyn2010majorana,PhysRevLett.105.177002}, one of the most widely discussed probes is the topological gap protocol~\cite{pikulin2021protocol,aghaee2023inas}, which uses tunneling conductance measurements to infer the closing and reopening of the bulk gap. However, it is a complex protocol requiring advanced statistical analysis~\cite{aghaee2023inas,legg2025commentinasalhybriddevices}. 
More generally, most existing probes rely on the bulk–boundary correspondence, in which boundary measurements are interpreted as signatures of a change in the bulk topological invariant~\cite{law2009majorana,PhysRevB.82.180516,2012PhRvL.109z7002L}. As a result, the bulk phase transition is typically inferred indirectly from edge phenomena rather than accessed through a direct bulk-sensitive observable. 
The issue with this approach is that it is also sensitive to non-topological physics at the boundary, for instance zero modes created by a smooth potential or quantum dot formation at the ends of the nanowire~\cite{PhysRevB.86.100503,
PhysRevB.96.075161,prada2020andreev}.  

In this work, we propose using the supercurrent as a probe of topological phase transitions, as a simple, bulk-sensitive, and experimentally accessible alternative to previous methods. 
A superconductor carries a supercurrent $j_q$ proportional to the Cooper-pair momentum $2q$ (we set $\hbar = 1$), $j_q=\rho_s q$, where the proportionality coefficient is the superfluid stiffness. In the limit of weak s-wave pairing and low temperature,
we can approximate the superfluid stiffness of a single band by $\rho_s=e n/m$, where $n$ is the electron density in the system, $m$ is the effective mass, and $e$ is the electron charge~\cite{tinkham2004introduction}. For a parabolic dispersion in one dimension, the ratio $n/m\propto v_F$, where $v_F$ is the Fermi velocity. More generally, in systems with multiple Fermi points, the superfluid stiffness is proportional to the sum of the Fermi velocities at a fixed chemical potential, $n/m\propto\sum_i k_{Fi}/m =\sum_i v_{Fi}$.

\begin{figure}[ht]
    \centering
    \includegraphics{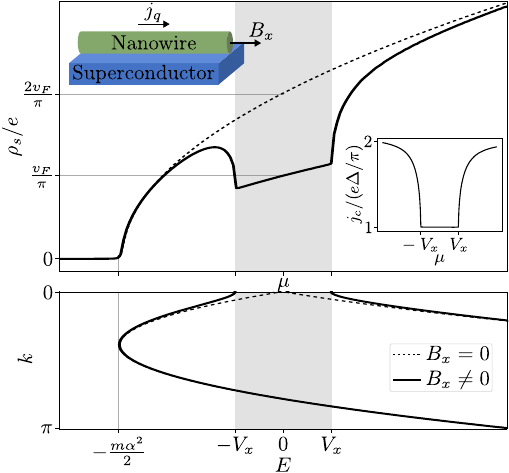}
\caption{
\textbf{Top:} Superfluid stiffness $\rho_s$ as a function of chemical potential $\mu$ for zero magnetic field ($B_x=0$, dashed lines) and finite magnetic field ($B_x\neq0$, solid lines). The top inset illustrates a nanowire proximitized by a superconductor carrying a supercurrent $j_q$ in the presence of a magnetic field ($ B_x$) applied along the wire. The bottom inset is the dimensionless critical current $j_c$ as a function of chemical potential $\mu$ near the topological gap, showing a suppression in the current for $|\mu|<V_x$.
\textbf{Bottom:} Normal-state Rashba dispersion for $B_x=0$ (dashed lines) and in the presence of a magnetic field $B_x\neq0$ (solid lines). The magnetic field opens a gap at $k=0$, whose magnitude corresponds to the suppression region of the superfluid stiffness in the top panel~\cite{large_soc}.
}
    \label{fig:fig1}
\end{figure}

The above argument suggests that in the topological phase, as one of the Fermi velocities goes to zero, the superfluid stiffness drops by a factor of 2 (from $2v_F/\pi$ to $v_F/\pi$)~\cite{large_soc}, somewhat similar to the behavior of normal state conductance in the helical gap~\cite{doi:10.1021/acs.nanolett.6b00051,Kammhuber_2017,doi:10.1021/acs.nanolett.7b00797,doi:10.1021/acs.nanolett.3c03604,pan2025ballistic}.  
Indeed, a careful calculation (detailed below) confirms this expectation; see Fig.~\ref{fig:fig1}. 
The non-monotonic dependence of $\rho_s$ on the chemical potential when the Zeeman field is applied serves as a measure of the bulk topological phase transition~\cite{large_soc}. Alternatively, one can measure the (depairing-limited) critical current $j_c=\rho_sq_c$, where $q_c$ is the critical momentum at which the energy gap for a quasiparticle excitation closes. In the inset of Fig.~\ref{fig:fig1}, we show that the critical current shows a similar non-monotonic behavior consistent with the topological phase transition.

There is a problem, however, with the above argument relying only on Fermi level properties:
One could also argue that since the density scales as $n \propto 2k_F$ and the effective mass remains constant, there should not be a non-monotonicity in $\rho_s$ as a function of chemical potential.
Thus, a simple Fermi-level picture alone does not provide a definitive conclusion about the behavior of the stiffness across the topological phase transition.

In this paper, we derive the superfluid stiffness as a function of chemical potential for a proximitized superconducting 
1D Rashba  nanowire, see Fig.~\ref{fig:fig1}. Remarkably, our results agree with the simplified picture, and we show that the superfluid stiffness is suppressed by a factor of 2 in the topological region, exhibiting non-monotonic behavior as a function of chemical potential (Fig.~\ref{fig:fig1}), consistent with the topological phase transition~\cite{large_soc}.

Furthermore, we clarify the above ``paradox'' 
by invoking a low-energy helical approximation near the Fermi level. Surprisingly, we show that in the absence of a magnetic gap and therefore band curvature, the helical states do not carry supercurrent. The applied field induces a negative band curvature in the lower helical band, thereby suppressing superfluid stiffness and explaining our findings.

\textbf{\emph{Model.~}}

We consider a one-dimensional nanowire proximitized by a superconductor (Fig.~\ref{fig:fig1}), described by the Hamiltonian $H=\frac{1}{2}\sum_{k}C_{k}^{\dagger}\mathcal{H}_\text{BdG}C_{k}$, written in terms of the Nambu spinor $C_{k}=(\begin{array}{cccc}
c_{k+q\uparrow} & c_{k+q\downarrow} & c_{-k+q\downarrow}^{\dagger} & -c_{-k+q\uparrow}^{\dagger}\end{array})^{T}$ where $c_{k\sigma}$ and $c^\dagger_{k\sigma}$ are the fermionic annihilation and creation operators at momentum $k$ and spin $\sigma=\uparrow, \downarrow$, respectively.
Here, the finite Cooper-pair momentum $2q$ corresponds to a pairing potential that, in the presence of supercurrent, acquires a modulation of the order parameter $\Delta(x)=\Delta e^{2iqx}$. A gauge transformation
$c_{k\sigma}\to e^{iqx}c_{k\sigma}$ 
removes the spatial phase from the pairing and shifts it into the kinetic term, yielding the momentum shifts appearing above.

We define the Bogoliubov-de Gennes (BdG) Hamiltonian as
\begin{align}
\label{eq:genericBdG}
    \mathcal{H}_{\text{BdG}} = \mathcal{H}_0+\frac{1}{2}\mu_{B}g\vec{B}\cdot\vec{\sigma}+\Delta\tau_{x},
\end{align}
where $\mathcal{H}_0$ is the kinetic Hamiltonian in the absence of magnetic field $\mathbf{B}$ and superconductivity given by the pairing potential $\Delta$. We define $\xi_{k\sigma}$ as the normal dispersion for $\sigma=\uparrow,\downarrow$, given by the eigenvalues of $\mathcal{H}_0$. Here, the Pauli matrices $\sigma_{xyz}$ and $\tau_{xyz}$ act on spin and particle-hole spaces, respectively. We also define the Zeeman energy $V_x=\frac{1}{2}\mu_{B}gB_x$.
The BdG eigenvector $\phi_{k,n}$ obeys
\begin{equation}\label{eq:H_k general}
\mathcal{H}_{\text{BdG}}\phi_{k,n}=E_{k,n}\phi_{k,n}, 
\end{equation}
where, in component form 
$\phi_{k,n}=\left(\begin{array}{cccc}
u_{k,n}^{+} & u_{k,n}^{-} & v_{k,n}^{+} & v_{k,n}^{-}\end{array}\right)^{T}$.
In \cite{SM}, we derive the expectation value for the current
\begin{equation}
\label{eq:current}
    j_q =\frac{e}{L}\sum_{kn}(\partial_{k}\xi_{k+q\uparrow}|u_{k,n}^{+}|^{2}+\partial_{k}\xi_{k+q\downarrow}|u_{k,n}^{-}|^{2})f(E_{k,n}),
\end{equation}
where $L$ is the wire length, taken to infinity later.

For a parabolic dispersion in the absence of a magnetic field and spin-orbit coupling, the Hamiltonian from Eq.~\eqref{eq:genericBdG} can be diagonalized exactly, as shown in \cite{SM}. 
In the limit of zero temperature and weak pairing, the supercurrent for small $q$ is $j_q=en_{s}q/m$, where $n_{s}=2k_{F}/\pi+\mathcal{O}(\Delta^{2})$. 
Here we consider Cooper pair momentum smaller than the critical one, $q<q_c$,  where $q_c=\Delta/v_F$ is the value at which the quasiparticle gap closes, and superconductivity is destroyed. 

\textbf{\emph{Rashba nanowire.~}}

\begin{figure}
    \centering
    \includegraphics{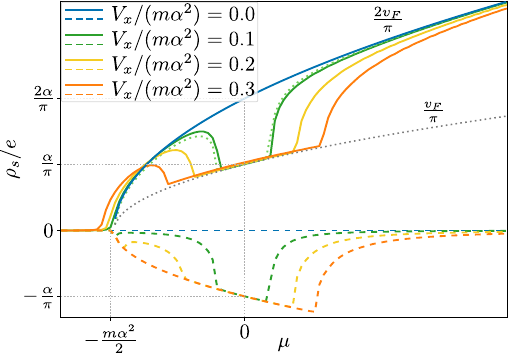}
    \caption{Superfluid stiffness $\rho_s/e$ as a function of chemical potential $\mu$ for different values of magnetic field applied in units of $V_x/(m\alpha^2)$. The continuous lines refer to the superfluid stiffness obtained from the Rashba Hamiltonian, Eq.~\eqref{eq:rashba}, while the dashed lines are obtained from a linear approximation around the Fermi level, i.e., the helical Hamiltonian in Eq.~\eqref{eq:helical}. For $V_x/(m\alpha^2)=0.1$, we also show the parabolic $\rho_s/e$ in the absence of a magnetic gap ($2v_F/\pi$) added to the stiffness from the linearized system (dotted green), which matches the suppressed superfluid stiffness from the full Rashba 
    Hamiltonian (continuous green). The suppressed current for all values of applied field aligns with half the value of the total supercurrent (dotted gray), corresponding to $v_F/\pi$. }
    \label{fig:fig2}
\end{figure}

Now we consider the Rashba spin-orbit coupling, with the kinetic  Hamiltonian in Eq.~\eqref{eq:genericBdG} defined as 
\begin{equation}
\mathcal{H}^R_0=\left(\frac{k^{2}+q^{2}}{2m}-\mu\right)\tau_{z}+\frac{kq}{m}+\alpha\left(k\tau_{z}+q\right)\sigma_{z},
\label{eq:rashba}
\end{equation}
where $\alpha$ is the Rashba spin-orbit coupling, which will spin-split the two parabolic bands, separating the Fermi momenta $k_F=\pm\alpha m +\sqrt{2m\mu+m^2\alpha^2}$. With an applied magnetic field $B_x$ along the wire, the spin-up and spin-down bands hybridize, creating a topological gap in the spectrum; see Fig.~\ref{fig:fig1}. 
Throughout, we focus on the spin-orbit-dominated regime $\Delta<V_x<m\alpha^2$.

We diagonalize the Hamiltonian of Eqs.~\eqref{eq:genericBdG} and~\eqref{eq:rashba} numerically and calculate the supercurrent from Eq.~\eqref{eq:current}.
We show in Fig.~\ref{fig:fig2} the superfluid stiffness as a function of chemical potential $\mu$ for various values of magnetic field. 
For zero applied field $V_x=0$, the superfluid stiffness 
is equal to the parabolic result, as there is no topological gap in the system. 
 For non-zero $V_x$, the supercurrent is suppressed inside the topological gap by half the amount without a magnetic field, 
 shown by the gray dotted line in Fig.~\ref{fig:fig2}. As we increase the strength of the magnetic field, the region where the current is suppressed increases as $2V_x$, which is the size of the topological gap. However, the amount of suppression is not dependent on $V_x$; it is always half of the zero-field value in the weak-pairing limit $\Delta\ll V_x$. 
 
\textbf{\emph{Interpretation and helical model.~}}

 To understand the mechanism that suppresses the supercurrent, we use a low-energy model consisting of one helical band described by a kinetic Hamiltonian 
 \begin{equation}
     \label{eq:helical}
     \mathcal{H}^H_0=v_{F}\left(k\sigma_{z}-k_{F}\right)\tau_{z}+v_{F}q\sigma_{z},
 \end{equation} in Eq.~\eqref{eq:genericBdG}. This single helical band is obtained by linearizing around the inner Fermi momentum $k_F$, which is strongly influenced by the topological gap.

In the absence of a Zeeman field, the BdG Hamiltonian commutes with $\sigma_z$, and the system can be solved analytically. 
The quasiparticle energies for the helical Hamiltonian were calculated in ~\cite{de2023superconducting}, and for the bottom branches, we have 
\begin{align}
E_{k,\sigma} & =\sigma qv_{F}-\sqrt{\Delta^{2}+ \left(k - \sigma k_{F}\right)^{2}  v_{F}^{2}}.
\label{eq:spectrum-toy-model}
\end{align}
where $\sigma=+,-=\uparrow,\downarrow$. The amplitudes for the bottom bands are given by [in the notation of Eq.~\eqref{eq:H_k general}]
    \begin{align}
        \label{eq:amplitudes_helical}
        \left|u_{k,\downarrow}^{-}\right|^{2}&=\frac{1}{2}\left[\frac{v_{F}(k+k_{F})}{\sqrt{v_{F}^{2}(k+k_{F})^{2}+\Delta^{2}}}+1\right],\\
        \left|u_{k,\uparrow}^{+}\right|^{2}&=\frac{1}{2}\left[\frac{-v_{F}(k-k_{F})}{\sqrt{v_{F}^{2}(k-k_{F})^{2}+\Delta^{2}}}+1\right],\\
        \left|u_{k,\downarrow}^{+}\right|^{2}&=\left|u_{k,\uparrow}^{-}\right|^{2}=0.
    \end{align}

For the linear dispersion, the partial derivatives in the current, Eq.~\eqref{eq:current}, are equal to the Fermi velocities in the system, $\partial_k\xi_{k,\sigma}=\sigma v_F$. 
For $q<q_c$, the quasiparticle spectrum, Eq.~\eqref{eq:spectrum-toy-model}, remains gapped. In the low-temperature limit compared to this gap, the Fermi-Dirac distribution in Eq.~\eqref{eq:current} vanishes for positive energies and equals one for negative energies. Using the amplitudes from Eq.~\eqref{eq:amplitudes_helical}, the current then reduces to a sum of the odd function $|u_{k,\uparrow}^{+}|^{2} - |u_{k,\downarrow}^{-}|^{2} $ over a symmetric interval, which vanishes by symmetry. Therefore, $j_q=0$, implying that the helical states do not carry supercurrent in the absence of a magnetic field.

However, when a magnetic field is applied to the system, the opposite-spin species in the helical model mix, opening a helical gap at $k=0$. Therefore, the spectrum becomes curved. This curvature enables the appearance of a supercurrent in the system. 

This is most easily seen in the $k_F=0$ case, where the lower band is fully filled, and the superfluid stiffness can be found analytically. 
Although the presence of a magnetic gap does not affect the velocities $\partial_k\xi_{k\sigma}$ in Eq.~\eqref{eq:current}, it alters the amplitudes, which now acquire a more complicated form than Eq.~\eqref{eq:amplitudes_helical}. For small $q$, where the current is proportional to the Cooper pair momentum, we can obtain the superfluid stiffness as
\begin{align}
    \label{eq:integrand}
    \rho_s = \partial_q j_q|_{q=0} &=ev^2_F\int_{-\infty}^{\infty}\frac{dk}{2\pi}\left[ \frac{v_F^2k^2-\left( V_x-\Delta \right)\Delta}{2V_x\Delta\sqrt{v_F^2k^2+\left( V_x-\Delta \right)^2}} \right. \nonumber \\& \left. - \frac{v_F^2k^2+\left( V_x+\Delta \right)\Delta}{2V_x\Delta\sqrt{v_F^2k^2+\left( V_x+\Delta \right)^2}}  \right] .
\end{align}
Integrating over all $k$ in Eq.~\eqref{eq:integrand}, we find $\rho_s =-ev_F/\pi + \mathcal{O}(\Delta^2)$, in the zero-temperature and weak-pairing limit. 

In Fig.~\ref{fig:fig2} we show the superfluid stiffness as a function of chemical potential for the helical Hamiltonian that linearizes the parabolic Rashba from Eq.~\eqref{eq:rashba}, such that $v_F=\sqrt{2\mu/m+\alpha^2}$. For chemical potential far below the helical gap, the dispersion is approximately linear, and the helical states do not carry supercurrent, as the system is far from the magnetic gap. As the chemical potential increases, the curvature induced by the magnetic gap yields a negative contribution to the supercurrent, matching the suppression of the supercurrent at the topological gap observed in the gapped Rashba system. 

The negative contribution to the supercurrent originates from an effective mass generated by the magnetic gap. To see this, consider the helical Hamiltonian in the absence of superconductivity, written in the $2\times2$ form
\begin{equation}
\mathcal{H}_\text{helical}=v_Fk\sigma_z-\mu+V_x\sigma_x.
\end{equation}
When $\mu\leq0$, the upper band is empty, and creating excitations in this band requires a large energy of at least $V_x$. In this limit, the system can be projected onto the bottom band as
\begin{equation}
H \approx \sum_k \epsilon_k\Gamma_{k}^{\dagger}\Gamma_{k},
\end{equation}
with $\epsilon_k=-\mu-\sqrt{v_F^2k^2+V_x^2}$. For $\mu < -|V_x|$, the bottom band is partially filled. 
The projected electron operators are 
\begin{equation}
c_{k\uparrow} \approx 
-\Gamma_{k} \sin{\frac{\theta}{2}} , \,\,
c_{k\downarrow} \approx 
\Gamma_{k} \cos{\frac{\theta}{2}},
\label{eq:operators_c_Gamma}
\end{equation}
where $\theta=\arccos{\left(v_Fk/\sqrt{v_F^2k^2+V_x^2} \right)}$. 

\begin{figure}
    \centering
    \includegraphics{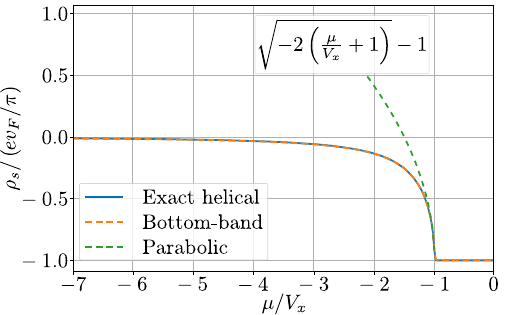}
    \caption{Dimensionless superfluid stiffness in units of $ev_F/\pi$ for a constant $v_F$ as a function of $\mu/V_x$. The exact helical (blue) and bottom-band projections (dashed orange) match for $\mu/V_x<0$, remaining constant and negative for $|\mu/V_x|<1$, implying that the helical bands carry a negative supercurrent due to the topological gap. In dashed green, we show the superfluid stiffness when the bottom helical band is approximated by a parabolic spectrum with negative effective mass $m_e = -V_x/v_F^2$.}
    \label{fig:fig3}
\end{figure}

Consider now the system with a pairing potential in the mean-field approximation. Using the bottom-band projection and the small-momentum approximation, the superconducting pairing term becomes
\begin{equation}
H_{\Delta}=\Delta c_{k+q\downarrow}c_{-k+q\uparrow}+h.c.\approx \frac{\Delta}{2}\Gamma_{k+q}\Gamma_{-k+q}+h.c.,
\end{equation}
where we took the limit of small momentum $k \ll V_x /v_F$, so that  the angle $\theta $ in Eq.~\eqref{eq:operators_c_Gamma} becomes $\pi/2$ and the operators take a simpler form. 
Introducing the Nambu spinor $\Psi_k=\begin{pmatrix}\Gamma_{k+q}, \Gamma_{-k+q}^{\dagger}\end{pmatrix}$, the BdG Hamiltonian for the bottom band projection becomes
\begin{equation}
\label{eq:bottom_hamiltonian}
\mathcal{H_{\text{BdG}}^\text{bottom}}=\left(\begin{array}{cc}
\epsilon_{k+q} & \Delta\\
\Delta & -\epsilon_{-k+q}
\end{array}\right).
\end{equation}
Diagonalizing this Hamiltonian, we write 
\begin{equation}
\label{eq:dandf}
\left(\begin{array}{c}
\Gamma_{k+q}\\
\Gamma_{-k+q}^{\dagger}
\end{array}\right)=\sum_{n=\pm}\left(\begin{array}{c}
d_{k,n}\\
f_{k,n}
\end{array}\right)\Lambda_{k,n},
\end{equation}
where $\Lambda_{k,n}$ and $\Lambda_{k,n}^\dagger$ are the superconducting creation and annihilation operators and $d_{k,n}$ and $f_{k,n}$ are amplitudes solved in \cite{SM}. With the current operator $j	=\frac{e}{L}\sum_{k\sigma}\partial_{k}\xi_{k\sigma}c_{k\sigma}^{\dagger}c_{k\sigma}$, derived in \cite{SM}, and the definition from Eqs.~\eqref{eq:operators_c_Gamma} and~\eqref{eq:dandf}, the supercurrent becomes
\begin{equation}
    j_q =-\frac{e}{L}\sum_{k,n=\pm}\frac{v^2_{F}\left(k+q\right)}{\sqrt{\left(v_{F}k+v_{F}q\right)^{2}+V_{x}^{2}}}\left|d_{k,n }\right|^{2}f(E_{k,n})
    \label{eq:current_helical}
\end{equation}
By computing Eq.~\eqref{eq:current_helical}, we show in Fig.~\ref{fig:fig3} that the superfluid stiffness within the bottom-band projection matches the exact helical value (from Eq.~\eqref{eq:current}) for a constant Fermi velocity and a negative chemical potential.

Within the small-momentum approximation, expanding the bottom-band dispersion around $k=0$ yields $\epsilon_k\approx k^2/2m_e-\mu_e$, with effective mass $m_e=-V_x/v_F^2$ and chemical potential $\mu_e = \mu+V_x$. The magnetic gap, therefore, induces curvature in the otherwise linear helical spectrum, yielding a negative effective mass. In this case, Eq.~\eqref{eq:bottom_hamiltonian} reduces to a BdG Hamiltonian with a single parabolic band, which is solved in \cite{SM} and results in the superfluid stiffness proportional to the Fermi velocities. In Fig.~\ref{fig:fig3} we also show how the parabolic approximation with the negative effective mass is valid only near $\mu=-V_x$ and rapidly deviates from the exact helical result for larger values of $|\mu|$.

\textbf{\emph{Discussion and conclusions.~}}

In this work, we studied the supercurrent signature of a topological phase transition in a superconducting nanowire. 
A possible experimental realization consists of a semiconducting nanowire with strong Rashba spin–orbit coupling placed on top of a thin superconductor, which induces superconductivity in the wire via the proximity effect. An in-plane magnetic field applied along the wire opens a helical gap, while an electrostatic gate can be used to tune the system into and out of the topological phase. 
We showed that the bulk topological phase transition causes a non-monotonic behavior of the superfluid stiffness $\rho_s$ as a function of the chemical potential, providing a direct probe that does not rely on Majorana zero modes. In the spin-orbit-dominated regime, this non-monotonicity sets a necessary condition for the topological phase transition.

There are multiple ways to measure $\rho_s$. 
In a wire geometry, which hosts Majorana zero modes at its ends in the topological phase, the superfluid stiffness can be extracted from measurement of the kinetic inductance of the system~\cite{splitthoff2022gate,PhysRevResearch.6.043245}.
The bulk contribution to the kinetic inductance is extensive, unlike the Majorana contribution, allowing one to distinguish the two for a long wire. 
Similarly, this extensive contribution will be insensitive to the local physics near the end of the wire, such as smooth potential or quantum dot formation~\cite{PhysRevB.86.100503,
PhysRevB.96.075161,prada2020andreev}.

While the contribution to the superfluid stiffness from the parent superconductor is larger in magnitude, it is not expected to display the aforementioned features controlled by the chemical potential and can therefore be straightforwardly subtracted in experiments.
To make this concrete, we consider a realistic InAs nanowire proximitized by an Al layer, with parameters reported in Refs.~\cite{aghaee2023inas,microsoft2025interferometric}. In Ref.~\cite{SM}, we estimate the change in the kinetic inductance across the topological phase transition and find $\delta L\approx0.05$--$38$~pH, which falls well within the sensitivity of standard experimental techniques used to probe kinetic inductance~\cite{splitthoff2022gate,PhysRevResearch.6.043245,phan2022detecting}.

The non-monotonic signature requires $\Delta<V_x<m\alpha^2$, i.e., spin-orbit energies exceeding the induced gap, as expected in materials with strong spin-orbit coupling, such as InSb~\cite{Kammhuber_2017}, or small induced gap. In the opposite, Zeeman-dominated regime, $\rho_s$ is monotonic but shows characteristic kinks close to the topological phase transition that can be detected experimentally~\cite{SM}.

Alternatively, the topological phase transition can also be observed by measuring the (depairing-limited) critical current $j_c=\rho_s q_c$, which is suppressed in the topological phase (see inset in Fig.~\ref{fig:fig1}). 
It is worth noting that in our mean-field model, the critical momentum $q_c$ is determined from the depairing condition, defined as the value of $q$ at which the quasiparticle spectrum becomes gapless and superconductivity is destroyed. At the topological phase transitions at the chemical potentials $\mu=\pm\sqrt{V_x^2-\Delta^2}$, the system becomes gapless~\cite{PhysRevLett.105.177002,kurilovich2024nonlocal,murthy2020energy}, such that $q_c=0$. Consequently, at these points, the critical current obtained from the depairing condition is zero, and our theoretical model is not expected to be applicable. 
However, this does not affect our results in the neighboring gapped trivial and topological regions, where one has $q_c>0$ and 
a well-defined supercurrent response of the nanowire. The physics of the critical phase transition point~\cite{kurilovich2024nonlocal,murthy2020energy}, where our description breaks down, remains an interesting open direction.

\textbf{\emph{Acknowledgments.~}} 

We thank Jason Alicea, Valla Fatemi, Bernard van Heck,  Gustavo M. Monteiro, Dima Pesin, Ivana Petkovic, and Uri Vool for useful discussions. 
This work was supported by the U.S. Department of Energy, Office of Science, National Quantum Information Science Research Centers, Quantum Science Center
and  by the National Science Foundation award number DMR-2540313.  
I.W. was supported by the National Science Foundation through the Research Experience for Undergraduates program under Grant Number PHY-2244297.

\bibliography{refs}

\clearpage
\newpage

\setcounter{equation}{0}
\setcounter{figure}{0}
\setcounter{section}{0}
\setcounter{table}{0}
\setcounter{page}{1}
\makeatletter
\renewcommand{\theequation}{S\arabic{equation}}
\renewcommand{\thepage}{S-\arabic{page}}
\renewcommand{\thesection}{S\arabic{section}}
\renewcommand{\thefigure}{S\arabic{figure}}


\renewcommand{\thetable}{S-\Roman{table}}

\renewcommand\section{\@startsection{section}{1}{0pt}%
  {1.5ex plus 1ex minus .2ex}%
  {1ex plus .2ex}%
  {\centering\normalfont\bfseries}}
  

\begin{widetext}
\begin{center}
Supplementary materials on \\
\textbf{``Supercurrent as a bulk probe for topological phase''}\\

Tatiana de Picoli$^{1}$, Ian Wojtowicz$^{1,2}$, Jukka I. V{\"a}yrynen$^{1}$, \\ 
$^{1}$ \textit{Department of Physics and Astronomy, Purdue University, West Lafayette, Indiana 47907, USA} \\
$^{2}$ \textit{Department of Mechanical Engineering, University of Minnesota Twin Cities, Minneapolis, 55455, USA}

\end{center}

These supplementary materials contain details about how we obtain the current operator and its expectation value, which allow us to calculate the supercurrent and superfluid stiffness. We show the spin-degenerate BdG Hamiltonian and the expectation value for the parabolic dispersion analytically. Finally, we use realistic system parameters to estimate the kinetic inductance.

\section{Current operator and expectation value for supercurrent}
\label{sm:current_density}

The Bogoliubov de-Gennes (BdG) Hamiltonian describing a superconducting system $H=\frac{1}{2}\sum_{k}C_{k}^{\dagger}\mathcal{H}_\text{BdG}C_{k}$, written in terms of the Nambu spinor $C_{k}=\left(\begin{array}{cccc}
c_{k+q\uparrow} & c_{k+q\downarrow} & c_{-k+q\downarrow}^{\dagger} & -c_{-k+q\uparrow}^{\dagger}\end{array}\right)^{T}$ where $c_{k\sigma}$ and $c^\dagger_{k\sigma}$ are the annihilation and creation operators at momentum $k$ and spin $\sigma=\uparrow, \downarrow$, respectively.
Defining a new basis $C_{k}=\sum_{n}\phi_{k,n}\Lambda_{k,n}$, the Hamiltonian can be diagonalized as 
\begin{align}
    H=\frac{1}{2}\sum_{kn}E_{k,n}\Lambda_{k,n}^{\dagger}\Lambda_{k,n},
\end{align}
where $E_{k,n}$ and $\phi_{k,n}$ can be obtained from
\begin{equation}
\mathcal{H}_{\text{BdG}}\phi_{k,n}=E_{k,n}\phi_{k,n}. 
\end{equation}

The particle density operator in the momentum space is defined as
\begin{align}
    n_{p}&=\frac{1}{L}\sum_{k\sigma}c_{k\sigma}^{\dagger}c_{k+p\sigma},
\end{align}
where $L$ is the nanowire length and $p$ is a shift in the momentum.

To derive the current from the continuity equation, we use the Hamiltonian in the normal phase, this is, $H_{q,\Delta\rightarrow0}	=\sum_{k\sigma}\xi_{k\sigma}c_{k\sigma}^{\dagger}c_{k\sigma}$, neglecting both the pairing term, Cooper pair momentum, and the magnetic field.
The magnetic field term in the Hamiltonian commutes with the density operator, so it does not change the structure of the continuity equation itself, which follows from charge conservation. 
Both pairing and Cooper pair momentum can be neglected in the first moment due to the use of the BCS mean-field Hamiltonian, which does not conserve particle number. The resulting apparent source terms in the continuity equation are an artifact of the mean-field decoupling. As shown in ~\cite{PhysRevB.49.6841}, these source terms vanish in a self-consistent treatment, matching with the expected result for the interacting theory, which strictly conserves particle number.

Using the normal state Hamiltonian, with the Heisenberg equation $\left[H,n_{p}\right]=-i\frac{dn_p}{dt}$ and continuity equation $\frac{dn_{p}}{dt}=-ipj$, we can obtain the current operator as 
\begin{equation}
    j_{p}	
=\frac{e}{L}\sum_{k\sigma}\frac{\left(\xi_{k+p\sigma}-\xi_{k\sigma}\right)}{p}c_{k\sigma}^{\dagger}c_{k+p\sigma}, 
\end{equation}
which for the zero momentum $p=0$ becomes 
\begin{equation}
    j_{p\rightarrow0}  \equiv j	=\frac{e}{L}\sum_{k\sigma}\partial_{k}\xi_{k\sigma}c_{k\sigma}^{\dagger}c_{k\sigma}.
    \label{eq:current_operator}
\end{equation}
Since we are interested in the supercurrent, we can rewrite the creation and annihilation operators in terms of the operators $\Lambda _{k,n}$ and compute the expectation value of the supercurrent and density in the system by taking $\left\langle \Lambda_{k,n}^{\dagger}\Lambda_{k',n'}\right\rangle =\delta_{kk'}\delta_{nn'}f(E_{k,n})$, where $f(E_{k,n})$ is the Fermi-Dirac distribution at energy $E_{k,n}$ 
\begin{align}
\label{eq:appendixcurrent}
    \left\langle j\right\rangle \equiv j_q &=\frac{e}{L}\sum_{kn}\left(\partial_{k}\xi_{k+q\uparrow}\left|u_{k,n}^{+}\right|^{2}+\partial_{k}\xi_{k+q\downarrow}\left|u_{k,n}^{-}\right|^{2}\right)f(E_{k,n}),\\
    \label{eq:appendixdensity}
    \left\langle n_0\right\rangle \equiv n &=\frac{1}{L}\sum_{kn}\left(\left|u_{k,n}^{+}\right|^{2}+\left|u_{k,n}^{-}\right|^{2}\right)f(E_{k,n})
\end{align}

\section{Diagonalization and supercurrent for a spin-degenerate system}
\label{sm:parabolic_diagonalization}
The conventional superconducting Hamiltonian without spin-orbit coupling or Zeeman field can be written as
\begin{equation}
\label{Seq:bottom_hamiltonian}
\mathcal{H_{\text{BdG}}}=\left(\begin{array}{cc}
\epsilon_{k+q} & \Delta\\
\Delta & -\epsilon_{-k+q}
\end{array}\right),
\end{equation}
where we initially assume a generic dispersion $\epsilon_{k+q}$. We can define the average and difference of the dispersions as 
\begin{align}
    \bar{\epsilon}_k &= \frac{\epsilon_{k+q}+\epsilon_{-k+q}}{2}, \\
\delta\epsilon_k &= \frac{\epsilon_{k+q}-\epsilon_{-k+q}}{2},
\end{align}
respectively.
The eigenspinors of  Eq.~\eqref{Seq:bottom_hamiltonian} obey 
\begin{equation}
\mathcal{H}_{\text{BdG}}\left(\begin{array}{c}
d_{k,n}\\
f_{k,n}
\end{array}\right)=E_{k,n}\left(\begin{array}{c}
d_{k,n}\\
f_{k,n}
\end{array}\right), 
\end{equation}
where $d_{kn}$ and $f_{kn}$ are amplitudes and the energies are $E_{k,\pm}=\delta\epsilon_k \pm \sqrt{\Delta^2+\bar{\epsilon}_k^{\,2}}$.
Here we omit the spin index, assuming that the spin bands are degenerate.
Diagonalizing the Hamiltonian for a generic dispersion, we find the squared amplitudes  given by 

\begin{equation}
\label{eq:ap_amplitudes_squared}
    \left|d_{k,\pm}\right|^{2}	=\frac{1}{2}\left[1\pm\frac{\bar{\epsilon}_k}{\sqrt{\Delta^{2}+\bar{\epsilon}_k^{2}}}\right],
\quad \left| f_{k,\pm}\right|^{2}	=\frac{1}{2}\left[1\mp\frac{\bar{\epsilon}_k}{\sqrt{\Delta^{2}+\bar{\epsilon}_k^{2}}}\right].
\end{equation}

For the particular case of a parabolic dispersion without Zeeman field and spin-orbit coupling, the Hamiltonian of Eq.~\eqref{eq:genericBdG} is block diagonal, with each of the blocks identical since the spin species are degenerate; therefore, we find 
\begin{align}
\label{eq:ap_amplitudes_parabolic}
    u^+_{k,n}&=u^-_{k,n}=d_{k,n},\\
    v^+_{k,n}&=v^-_{k,n}=f_{k,n}, \quad n = \pm,
\end{align}
where the average $\bar{\epsilon}_k$ takes the value of $\bar{\epsilon}_k\rightarrow \frac{k^{2}+q^{2}}{2m}-\mu$ for the parabolic dispersion.
Note that the average $\bar{\epsilon}_k$ and the squared amplitudes, Eq.~\eqref{eq:ap_amplitudes_squared}, are even in $k$ (and $q$ for parabolic dispersion), which will be relevant for the calculation of current in the next section.

\subsection{Supercurrent for parabolic dispersion}

Here we derive the current expectation value for a spin-degenerate system with parabolic dispersion analytically. In the zero-temperature limit and taking the continuous limit of Eq.~\eqref{eq:appendixcurrent}, the current becomes
\begin{equation}
\label{eq:ap_jparab}
     j_q =\frac{e}{\pi }\int_{-\infty}^{\infty} dk  \frac{k+q}{m}\left(\left|d_{k,-}\right|^{2} f(E_{k,-}) + \left|d_{k,+}\right|^{2} f(E_{k,+}) \right)
\end{equation}
where we used the amplitudes from Eq.~\eqref{eq:ap_amplitudes_parabolic} and energies $E_{k,\pm}=\frac{kq}{m} \pm \sqrt{\Delta^2+\bar{\epsilon}_k^{\,2}}$.

Let us consider the limit $\Delta \to 0$, keeping $q$ non-zero. In this limit, the amplitudes $d_{k,\pm}$ become sign functions of $k$, while at zero temperature, the Fermi functions become step functions. 
One can show that the integral and the current vanish in this case. 
A much easier way to see this is to use a different set of eigenstates obtained directly from S8 with $\Delta = 0$. Only the eigenstate with energy $\epsilon_{k+q}$ has a non-vanishing amplitude $d_k = 1$ (while $f_k = 0$), valid for all $k$. 
The integrand in $j_q$ is then just $(k+q) f(\epsilon_{k+q}) /m$, which, with a change of variables from $k\rightarrow k+q$, becomes an odd integrand over an even interval and vanishes upon integration. 
Thus, in the absence of superconductivity, Eq.~\eqref{eq:ap_jparab} yields the expected, vanishing supercurrent.

For finite pairing $\Delta$, and Cooper pair momentum below the critical momentum $q\leq q_c$, the energy spectrum is gapped; therefore, in the zero-temperature limit, the Fermi functions will take the values of $f(E_{k,+})=0$ and $f(E_{k,-})=1$ for all $k$. Since the amplitude $\left|d_{k,-}\right|^{2}$ only depends on $k$ through the average $\bar{\epsilon}$ and the average is even in $k$, the term proportional to $k/m$ in Eq.~\eqref{eq:ap_jparab} vanishes due to symmetry.
The only term that survives is the one proportional to the Cooper pair momentum $q$. Using the density definition from Eq.~\eqref{eq:appendixdensity}, the current can be written as
\begin{equation} \label{app:eq:jq}
    j_q =\frac{q}{m}\frac{e}{\pi }\int_{-\infty}^{\infty} dk\left|d_{k,-}\right|^{2}= \frac{q}{m}en,
\end{equation}
where $n$ is the expectation value of the total electron density in the system and depends on $\Delta$ and $q$. 

Furthermore, in the weak-pairing limit with $q\ll q_c$, the supercurrent can be written in terms of the normal-state density $n_s=2k_F/\pi$, which is independent of the pairing amplitude. In the limit of $\Delta\rightarrow 0$, the amplitude $\left|d_{k,-}\right|^{2}$ becomes $(1-\operatorname{sign}(\bar{\epsilon}_k))/2$, which does not lead to a vanishing supercurrent anymore. This behavior seems contradictory compared to our earlier result of obtaining zero supercurrent when considering $\Delta=0$ from the beginning in comparison to taking $\Delta\rightarrow 0$ later on. The resolution is that Eq.~\eqref{app:eq:jq} assumes $q \leq q_c$, which cannot be satisfied by keeping $q$ fixed and taking $\Delta = 0$.  

\section{Kinetic inductance estimates for a realistic device}
\label{sm:inductance_largeVx}

The superfluid stiffness can be inferred from the system's kinetic
inductance. We consider an InAs nanowire proximitized by an Al layer, with parameters
similar to those reported in Ref.~\cite{microsoft2025interferometric},
namely length $l=3\,\mu\text{m}$, width $w=60\,\text{nm}$, and thickness
$t_{\text{Al}}=3.5\,\text{nm}$ for the Al layer. The kinetic inductances for the Al layer and
the InAs nanowire are given by 
\begin{align}
L_{\text{Al}} & =\mu_{0}\lambda_{L}^{2}\left(\frac{l}{wt_{\text{Al}}}\right),\\
L_{\text{InAs}} & =\frac{hl}{4e^{2}v_{F}},
\end{align}
where $\mu_{0}$ is the vacuum permeability and $\lambda_{L}$ is the
penetration depth in the thin Al layer. In the nanowire, $L_{\text{InAs}}$ is the ballistic kinetic inductance just outside the topological phase, written here for two channels with Fermi velocity $v_F$; for channels with different velocities $v_1$ and $v_2$, $2v_F\rightarrow v_1+v_2$. 
For the nanowire and the superconductor in parallel, $1/L_{\text{total}}=1/L_{\text{InAs}}+1/L_{\text{Al}}$.

We consider measurements at a fixed magnetic field, in which a gate tunes the chemical potential across the topological transition. At fixed $B$, $L_{\text{Al}}$ remains fixed while $L_{\text{InAs}}\rightarrow L'_{\text{InAs}}=\eta L_{\text{InAs}}$ as $\mu$ is tuned into the topological phase, where $\eta=\rho_s(\mu_{\text{out}})/\rho_s(\mu_{\text{in}})$ is the ratio of the nanowire superfluid stiffness just outside and inside the topological phase.
When $L_{\text{InAs}}\gg L_{\text{Al}}$, the change in the inductance can be approximated by
\begin{equation}
\delta L\approx L_{\text{Al}}^{2}\left(\frac{1}{L_{\text{InAs}}}-\frac{1}{L'_{\text{InAs}}}\right)=\left(1-\frac{1}{\eta}\right)\frac{L_{\text{Al}}^{2}}{L_{\text{InAs}}}.\label{eq:deltaL}
\end{equation}
The penetration depth depends sensitively on the layer thickness~\cite{lopez2025superconducting} and, to our knowledge, is not known for our considered $t_{\text{Al}}$. Assuming that $\lambda_{L}$
is between $150$ and $600\,\text{nm}$~\cite{lopez2025superconducting},
we find $L_{\text{Al}}$ between $0.40$ and $6.5\,\text{nH}$.

For
the nanowire, $m\approx0.032\,m_{e}$, $\alpha\approx8.4$~meV\,nm~\cite{aghaee2023inas}, and $\Delta\approx0.11$~meV~\cite{microsoft2025interferometric}.
At $B\approx2$~T, where the measurements of Ref.~\cite{microsoft2025interferometric}
were performed, $V_{x}\approx0.26$~meV, using the renormalized g-factor $|g_{\text{eff}}|\approx4.5$ from Ref.~\cite{aghaee2023inas},
while $m\alpha^{2}\approx0.03$~meV, corresponding to a system in the Zeeman-dominated regime, i.e.,  $V_{x}\gg m\alpha^{2}$, see~\ref{sm:zeeman}. However, to find the topological and non-monotonic behavior discussed in the main text, the bottom band must have a local maximum at $k=0$, which requires $\Delta<V_{x}<m\alpha^{2}$, which we call the spin-orbit-dominated regime.

\subsection{Spin-orbit-dominated regime}
\label{sm:SOC}

Consider a case where the induced gap and the Zeeman energy are smaller than the spin-orbit energy. For the same $\alpha$, this requires a weaker induced gap, for example $\Delta\approx5~\mu$eV and $V_x\approx10~\mu$eV ($B\approx0.08$~T). In this case, inside the topological phase only the outer pair of Fermi points remains, while just outside it a second pair is present. Both have the velocity $v_{F}\approx\alpha/\hbar$, resulting in $\eta\approx2$. Using $v_{F}\approx1.3\times10^{4}$~m/s, we find
$L_{\text{InAs}}\approx1.5~\mu$H, yielding $\delta L\approx(1-1/2)L_{\text{Al}}^{2}/L_{\text{InAs}}\approx\left(0.05\sim14\right)$~pH.

\subsection{Zeeman-dominated regime}
\label{sm:zeeman}
Using the parameters reported in Refs.~\cite{aghaee2023inas,microsoft2025interferometric}, the system is in the Zeeman-dominated regime, i.e.,  $V_{x}\gg m\alpha^{2}$.
In this regime, the bands are spin-polarized parabolas, $\xi_{k,\pm}\approx k^{2}/2m\pm V_{x}$,
with Fermi velocities $v_{\pm}=\sqrt{2(\mu\mp V_{x})/m}$. This results in a monotonic
superfluid stiffness with a kink at the transition $\mu=V_{x}$, where the top band
starts to be filled; see Fig.~\ref{figsm:zeeman_dom}.

Just inside the topological phase, at $\mu=V_x$, only the bottom band contributes, with velocity
$v_{-}=2\sqrt{V_{x}/m}\approx7.6\times10^{4}$~m/s, for
$V_{x}\approx0.26$~meV~\cite{aghaee2023inas,microsoft2025interferometric}. Just outside, at $\mu=V_x+\delta\mu$, the top band adds a pair of Fermi points with velocity $v_{+}=\sqrt{2\delta\mu/m}$. Since pairing smears the kink over $\delta\mu\sim\Delta$, we take $\delta\mu\approx\Delta$, which gives $v_{+}\approx3.5\times10^{4}$~m/s. Neglecting the smaller increase of $v_{-}$ over this window ($\approx0.8\times10^{4}$~m/s), we find $\eta\approx(v_{-}+v_{+})/v_{-}\approx1.46$ and
$L_{\text{InAs}}\approx0.35~\mu$H, yielding an inductance change
$\delta L\approx(1-1/\eta)L_{\text{Al}}^{2}/L_{\text{InAs}}\approx\left(0.1\sim38\right)$~pH.
Both estimates are accessible with existing kinetic-inductance measurements~\cite{splitthoff2022gate,PhysRevResearch.6.043245,phan2022detecting}.

\begin{figure}[h]
\centering \includegraphics[width=1\columnwidth]{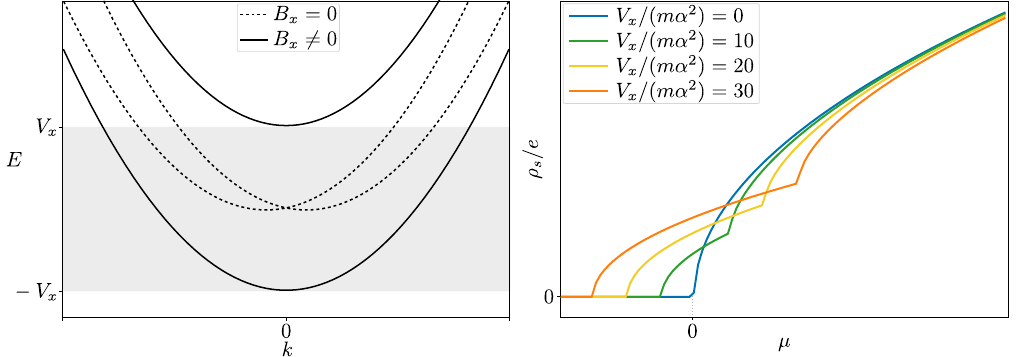}
\caption{(a) Normal-state dispersion in the Zeeman-dominated
regime, $V_{x}\gg m\alpha^{2}$. The shaded region marks
the topological phase, $|\mu|<V_{x}$. (b) Superfluid stiffness $\rho_{s}/e$
as a function of chemical potential $\mu$ in the Zeeman-dominated
regime.  }
\label{figsm:zeeman_dom} 
\end{figure}

\end{widetext}

\end{document}